\documentclass[11pt,showpacs,showkeys,aps,pra]{revtex4}
\usepackage[utf8x]{inputenc}
\usepackage{ucs}

\usepackage{amsmath,amssymb}
\usepackage{latexsym}
\usepackage{amsfonts}
\usepackage{dcolumn}
\usepackage{bm}
\usepackage[usenames]{color}
\usepackage{multirow}
\usepackage{graphicx}
\usepackage{hyperref}
\usepackage{subfig}
\usepackage{booktabs}
\usepackage{xcolor}
\usepackage[normalem]{ulem}
\usepackage{float}

\usepackage{float} 
\usepackage{wrapfig} 

\usepackage{upgreek} 
\usepackage{cancel} 
\usepackage{mathdots} 
\usepackage{mathrsfs} 
\graphicspath{{figures/}} 

\begin{document}


\title{Pauli effects in uncertainty relations}

\author{I.V. Toranzo$^{a,c}$, P. Sánchez-Moreno$^{b,c}$, R.O. Esquivel$^{c,d}$, J.S. Dehesa$^{a,c}$}
\affiliation{
$^a$Departamento de F\'isica At\'omica, Molecular y Nuclear, Universidad de Granada, 18071-Granada, Spain\\
$^b$Departamento de Matemática Aplicada II, Universidad de Granada, 18003-Granada, Spain\\
$^c$Instituto {\em Carlos I} de F\'isica Te\'orica y Computacional, Universidad de Granada, 18071-Granada, Spain\\
$^d$Departamento de Qu\'{\i}mica, Universidad Aut\'onoma Metropolitana, 09340-M\'exico D.F., M\'exico
}

\email{dehesa@ugr.es}
\date{\today}

\begin{abstract}
The Heisenberg uncertainty principle and Pauli exclusion principle are known to play a relevant role for a great diversity of quantum phenomena, ranging from the determination of quantum states till the stability of matter. However, although it is known that the Pauli principle qualitatively modifies the uncertainty relations, no explicit connection between these two fundamental pillars of quantum physics has yet been published in literature to the best of our knowledge. In this paper we analyze the effect of the Pauli principle in two mathematical formulations of the uncertainty principle: a generalised Heisenberg uncertainty relation valid for general fermion systems, and the Fisher-information-based uncertainty relation valid for fermion systems subject to central potentials. The accuracy of the Pauli-modified uncertainty relations is examined for all ground-state atoms from Hydrogen to Lawrencium in a numerically self-consistent method.
\end{abstract}



\maketitle

\section{Introduction}

The uncertainty principle and the Pauli principle are two fundamental pillars of quantum physics which have relevant consequences for the determination of quantum states of matter systems. Indeed, the former one prevents us to measure with arbitrary accuracy all the physical quantities which are classically within our reach, and the second one states that two identical fermions cannot occupy the same quantum state simultaneously. But perhaps the most distinguished issue of these two principles is the stability of matter \cite{lieb1}: atomic electrons operate as small radiating classical antennas that should fall on the nucleus at the time of a few billionths of a second, causing unstable atoms. The uncertainty principle comes to your rescue, enabling, together with the exclusion principle, the existence of electronic shells and subshells, and thus the periodic table and all the wealth of structural chemistry.\\
In fact, it is more than that. When we talk about the stability of microscopic systems (e.g., the stability of hydrogen), we simply mean that the total energy of the system cannot be arbitrarily negative. If the system would not have such lower bound to the energy, it would be posible to extract an infinite amount of energy, at least in principle. This stability of the first kind admits a generalization to the macroscopic systems, referred as stability of second kind. In this second type of stability, the lowest posible energy of the macroscopic systems depends at most linearly on the number of particles; or, in other terms, the lowest posible energy per particle cannot be arbitrarily negative as the number of particle increases. These two stability problems have a crucial relevance to understand the world around us. Both of them rely on the fermionic property of electrons; more specifically, they rely on the uncertainty principle and the Pauli principle.\\
The influence of the Pauli principle on the mathematical formulations of the uncertainty principle (i.e., the uncertainty relations) has been previously perceived (see e.g., \cite{basdevant}) but it has never been explicitly described, to the best of our knowledge. In this paper we tackle this issue. To be more specific, in our work we explore the effects of the Pauli exclusion principle on two concrete uncertainty relations of d-dimensional systems: a generalised Heisenberg relation valid for general finite fermion systems, and the Fisher-information-based uncertainty relation of systems moving in a central potential. In other words, we investigate the combined balance of the effects of spatial and spin dimensionalities on these two fundamental uncertainty relations. We do this way because of the relevant role that space dimensionality plays in the analysis of the structure and dynamics of natural systems and phenomena, from atomic and molecular physics (see e.g., \cite{herschbach,sen}), quantum optics \cite{rybin} to condensed matter (see e.g., \cite{acharyya,march,xluo}) and quantum information and computation (see e.g.,\cite{spengler,krenn}).\\
The structure of the present work is the following. In section II we first show the explicit dependence of a generalised Heisenberg uncertainty relation on the spin degree of freedom. As a particular case, the spin effects on the standard Heisenberg relation are given and the accuracy of the corresponding lower boun is examined for all atoms of the periodic table from Hydrogen to Lawrencium. In Section III, a similar study is carried out for the Fisher-information-based uncertainty relation in quantum systems with a central potential. Finally, some conclusions and open problems are given. Atomic units will be used throughout all the paper.

\section{Generalized Heisenberg uncertainty relations: Pauli effects}

Let us consider a $d$-dimensional system of $N$ independent fermions of spin $s$ moving in an arbitrary potential. Let us denote by $\rho(\vec{r})$ the position probability density of the system, whose moment around the origin or radial expectation value of order $\alpha$ is given by
\[
\langle r^\alpha \rangle  =   \int_{\mathbb{R}_{d}} \rho(\vec{r})\,r^{\alpha}\, d^{d}r, \, \, \alpha \ge 0,
\]
and $\langle p^2\rangle$ denote the corresponding radial momentum expectation value of second order. 
 In this section we find a lower bound to the uncertainty product $\langle r^\alpha \rangle^{\frac{2}{\alpha}}\langle p^2\rangle$,  $\alpha > 0$, which only depends on $d$, $N$ and $\alpha$. Then, for $\alpha = 2$ we obtain the standard Heisenberg uncertainty relation with the spin-dependent effects. Finally, we numerically analyze the accuracy of this relation for all atoms with nuclear charge $Z = 1$ through $103$.\\
First we show the main result: the generalised uncertainty relation of the form 
\begin{equation}
\label{eq:ineq1}
\langle r^\alpha \rangle^{\frac{2}{\alpha}}\langle p^2\rangle \geq A(\alpha, d) (2 s+1)^{-\frac{2}{d}} N^{\frac{2}{d}+\frac{2}{\alpha}+1},
\end{equation}
with the constant
\begin{equation}
\label{eq:cteA}
A(\alpha, d) = K(d) \cdot F(\alpha, d)\cdot C_{d}^{-\frac{2}{d}}\, ,
\end{equation}
where
\begin{equation}
\label{eq:cteK}
K(d) = \frac{4\pi d}{d+2}\left[\Gamma\left(\frac{d}{2}+1\right)\right]^{2/d},\\
\end{equation}
\begin{equation}
\label{eq:cteF}
F(\alpha, d) =\frac{2^{\frac{2}{d}+\frac{2}{\alpha}}\alpha^{1+\frac{4}{d}}(1+\frac{2}{d})^{1+\frac{2}{d}}}{\pi\left[\alpha(1+\frac{2}{d})+2\right]^{1+\frac{2}{d}+\frac{2}{\alpha}}}\cdot\left[\frac{\Gamma(\frac{d}{\alpha}+\frac{d}{2}+2)}{d(d+2)\Gamma(\frac{d}{\alpha})}\right]^{2/d},\\
\end{equation}
and  $1\leq C_d \leq 2$ for $d\geq 1$. There is a longstanding conjecture $C_d = 1$ due to Lieb and Thirring, which is now current belief; this is assumed heretoforth.\\
We will prove this result in two steps. First we use an inequality of Lieb-Thirring type \cite{lieb1} to bound the kinetic energy of the system in terms of the entropic moments
\[
W_{a}[\rho] = \int_{\mathbb{R}_{d}} \rho(\vec{r})^{a}\, d^{d}r, \, \, a \ge 1,
\]
and then, we bound these quantities in terms of the position moments $\langle r^{\alpha} \rangle$ of arbitrary order $\alpha$.
Indeed, the Lieb-Thirring inequality appropriately modified \cite{hundermartk} tells us that
\begin{equation}
\label{eq:inepe2}
	\langle p^{2}\rangle \geq K(d) (q\, C_{d})^{-\frac{2}{d}} W_{1+\frac{2}{d}}[\rho]\, , \quad \text{with} \quad q= 2s+1
\end{equation}
On the other hand, we can variationally bound the entropic moments $W_{a}[\rho]$ with the given constraints $\langle r^{0} \rangle = 1$ and $\langle r^{\alpha} \rangle$, $\alpha >0$. Following the lines of the method of Lagrange's multipliers described in Refs. \cite{dehesa88,dehesa89}, we obtain the lower bound
\begin{equation}
\label{eq:ineq2}
	W_{a}[\rho] \geq F(\alpha, a, d) \frac{N^{1+\frac{2}{d}+\frac{2}{\alpha}}}{\langle r^{\alpha} \rangle^{\frac{d}{\alpha}(a-1)}},
\end{equation}
with
\begin{equation}
\label{eq:cteF2}
	F(\alpha, a, d) = \frac{a^{a}\alpha^{2a-1}}{\left[ \Omega_{d}B\left(\frac{ad-d}{\alpha(a-1)},\frac{2a-1}{a-1}\right)\right]^{a-1}}\times \left\{ \frac{(ad-d)^{ad-d}}{[a(\alpha+ d)-d)]^{a(\alpha+ d)-d}}\right\}^{\frac{1}{\alpha}}
\end{equation}
and $\Omega_D = \frac{2\pi^{D/2}}{\Gamma(D/2)}$. The symbols $\Gamma(x)$ and $B(x,y) = \Gamma(x) \Gamma(y)/\Gamma(x+y)$ denote the well-known gamma and beta functions, respectively.\\
Now, putting $a = 1+\frac{2}{d}$ into expressions (\ref{eq:ineq2})-(\ref{eq:cteF2}) and multiplying the subsequent expression by the inequality (\ref{eq:inepe2}) one finally obtains the wanted generalised uncertainty relation (\ref{eq:ineq1}).

For the particular case $\alpha=2$, the generalized uncertainty relation (\ref{eq:ineq1}) gives the spin-dependent Heisenberg relation for $d$-dimensional $N$-fermion systems
\begin{equation}
\label{eq:ineq3}
\langle r^2 \rangle\langle p^2\rangle \geq A(2, d) (2 s+1)^{-\frac{2}{d}} N^{\frac{2}{d}+2}
\end{equation}
with 
\begin{equation}
\label{eq:cteA2}
A(2, d) = \left\{ \frac{d}{d+1}\left[ \Gamma(d+1)\right]^{1/d}\right\}^{2}.
\end{equation}
Let us note that the lower bound on the position-momentum Heisenberg product increases when both the spatial and spin dimensionalities are increasing, so that the uncertainty relation gets improved. Note that for large values of the spatial dimensionality $d$, the bound (\ref{eq:ineq3}) behaves as $d^{2}/e^{2}\, N^{2} = 0.13533 \, d^{2} N^{2}$. Since the general spinless $d$-dimensional bound is $d^{2}/4 \,N^{2} = 0.25 \, d^{2} N^{2}$, it is interesting to highlight that there is a delicate balance between the spatial and spin dimensionality effects so that it turns out that the lower bound (\ref{eq:ineq3}) is better or worse than the spinless bound when $d$ is small or large, respectively. This is because of the relative values of $d$, $s$, and $N$ in (\ref{eq:ineq3}).\\
 Moreover, for $d=3$ one trivially obtains the spin-dependent Heisenberg uncertainty relation for all real $N$-fermion systems
\begin{equation}
\label{eq:ineq4}
\langle r^2 \rangle\langle p^2\rangle \geq \left(\frac{3}{4}6^{1/3}\right)^{2} (2s+1)^{-2/3} N^{8/3}
\end{equation}
where the equality is reached for the harmonic oscillator, as previously known \cite{basdevant}.
\begin{center}
\begin{figure}[h]
\includegraphics[width=8cm]{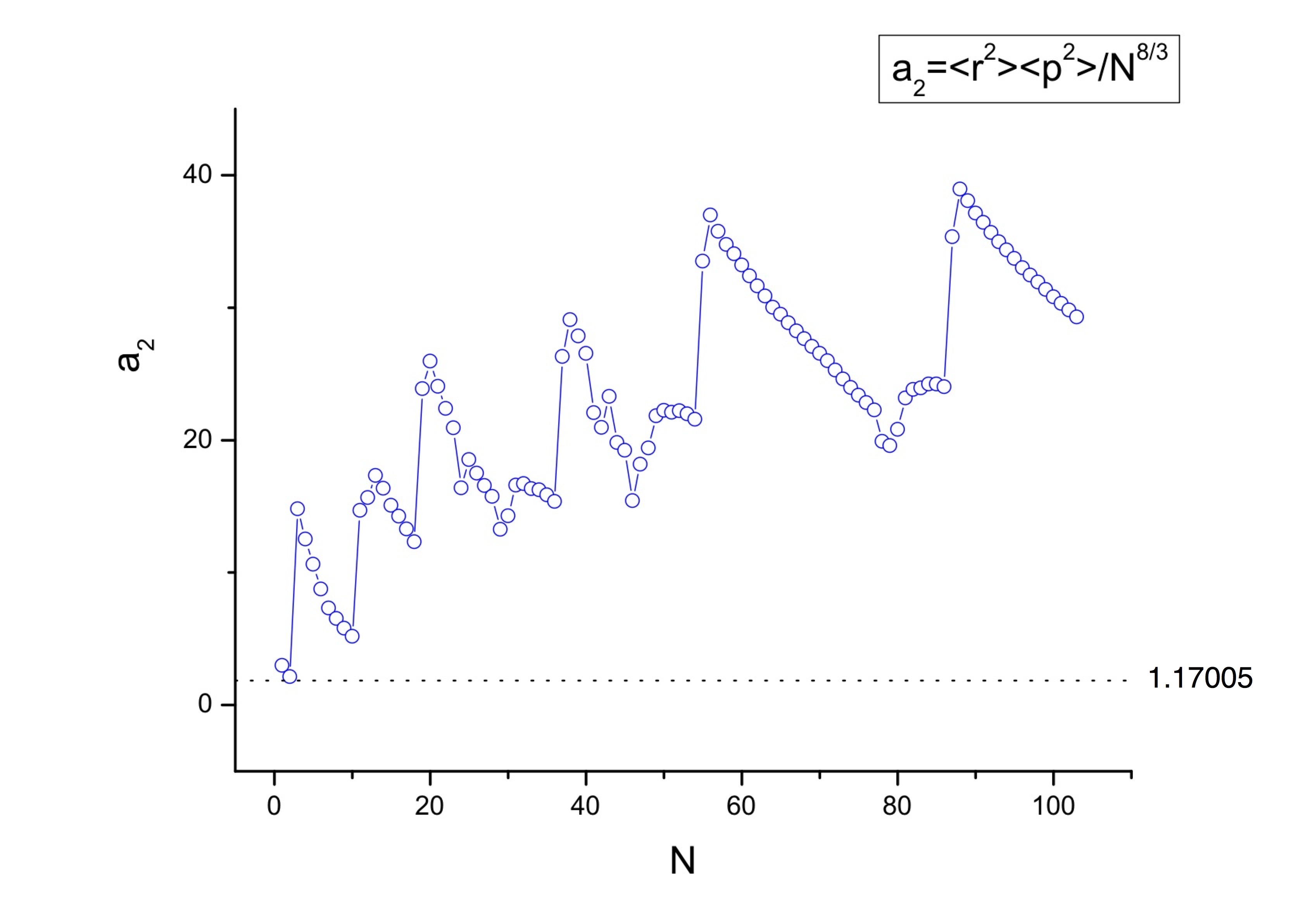}
\caption{Numerical study of the Heisenberg uncertainty relation for all neutral atoms with $N = 1$ to $103$.}
\label{fig:1}
\end{figure}
\end{center}
To have an idea of the accuracy of the Heisenberg relation (\ref{eq:ineq4}), we study the uncertainty product $\langle r^2 \rangle\langle p^2\rangle/N^{8/3}$ for all neutral atoms of the periodic table from Hydrogen ($N = 1$) to Lawrencium ($N = 103$) in a numerical Hartree-Fock framework. The results are shown in Fig. \ref{fig:1}, where the lower bound is $\left(\frac{3}{4}6^{1/3}\right)^{2} 2^{-2/3} = 1.17005$ since the electron spin $s = 1/2$. Therein we can clearly observe two important phenomena: (i) the atomic shell structure is grasped by the Heisenberg uncertainty relation, and (ii) the accuracy of the inequality globally decreases when the nuclear charge of the atoms is increasing.

\section{Fisher-information-based uncertainty relation: Pauli effects}

\begin{center}
\begin{figure}
\includegraphics[width=8cm]{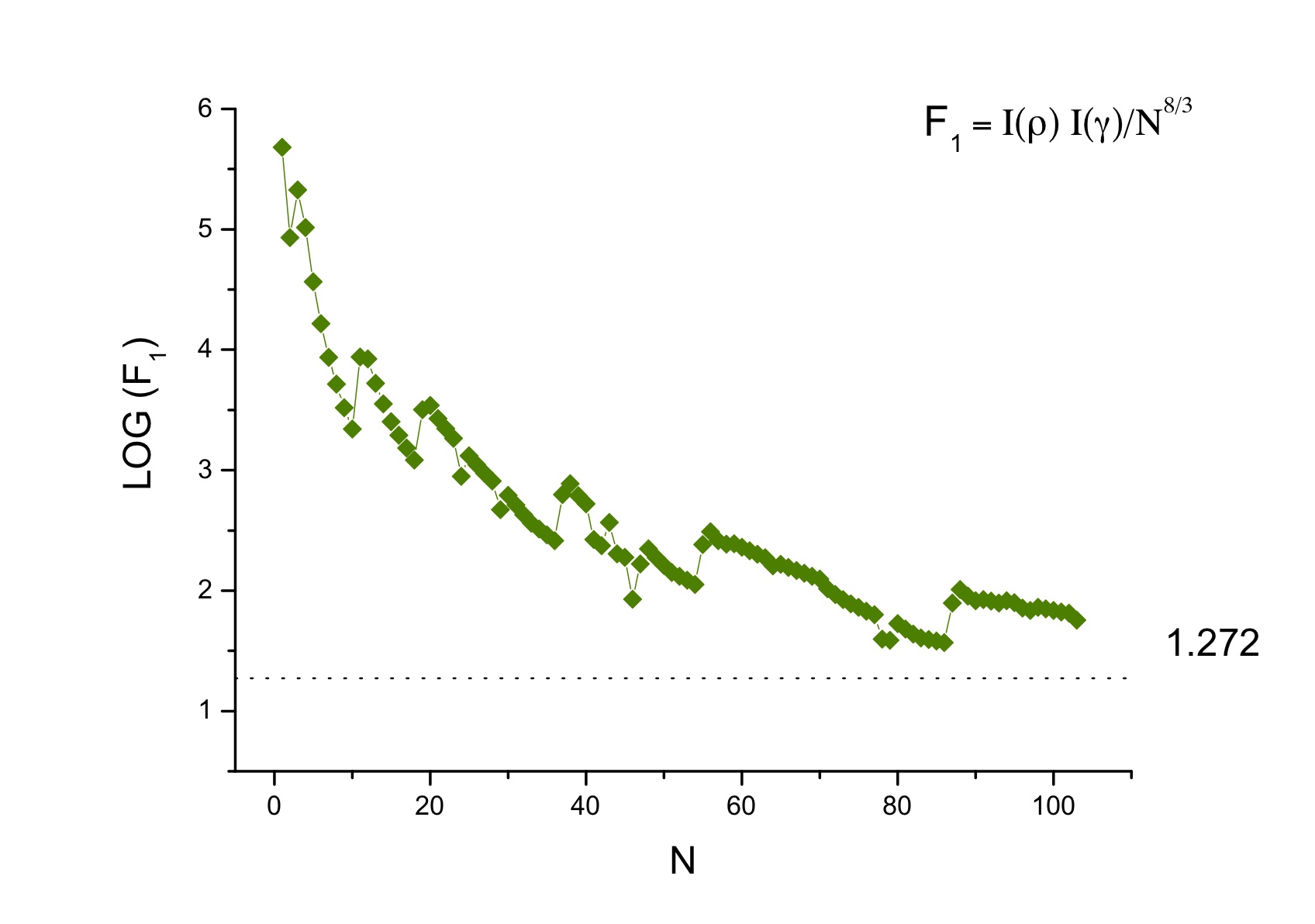}
\caption{Numerical study of the Fisher-information-based uncertainty relation for all neutral atoms with $N = 1$ to $103$.}
\label{fig:2}
\end{figure}
\end{center}

The Fisher information of a $d$-dimensional system of $N$ fermions characterized by the quantum-mechanical probability densities $\rho(\vec{r})$ and $\gamma(\vec{p})$ in position and momentum spaces are defined \cite{frieden_04,fisher} by  
\[
I[\rho] := \int \frac{|\vec{\nabla} \rho(\vec{r})|^{2}}{\rho(\vec{r})}\, d^{d} r \quad \text{and} \quad  I[\gamma] := \int \frac{|\vec{\nabla} \gamma(\vec{p})|^{2}}{\gamma(\vec{p})}\, d^{d} p.
\]
It has been proved that the product of these information-theoretical quantities is known to have an uncertainty character \cite{rom05,rom06,dehesa07}. In fact, for the systems subject to central potentials of arbitrary type one knows that the Fisher-information-based uncertainty product is bounded from below by the standard Heisenberg uncertainty product \cite{dehesa07} as
    \[
	I[\rho]\times I[\gamma] \geq 16 \left( 1- \frac{2|m|}{2l+d-2}\right)^2 \langle r^{2}\rangle\langle p^{2}\rangle,
	\] 
	where $l = 0,1,2,...$ and $m$ are two hyperquantum numbers. We should keep in mind that the angular part of the wavefunctions of a particle in a $d$-dimensional central potential is characterized by $d-1$ hyperangular quantum numbers $l\equiv \mu_1 \geq \mu_2 \geq \dots \geq \mu_{d-1}=\left| m \right| \geq 0$.
\\   
Then, taking into account this inequality together with the expressions (\ref{eq:ineq3})-(\ref{eq:cteA2}) one finds the following spin-modified Fisher-information-based uncertainty relation
\begin{equation}
\label{eq:ineq5}
I[\rho] \times  I[\gamma] \geq C(l,m,d) (2 s+1)^{-\frac{2}{d}} N^{\frac{2}{d}+2},
\end{equation}
with the constant
\begin{equation}
\label{eq:cteC}
C(l,m,d) = 16 \left[1 - \frac{2|m|}{2l+d-2}\right]^{2}\times \left\{\frac{d}{d+1}[\Gamma(d+1)]^{\frac{1}{d}}  \right\}^{2}.
\end{equation}
This uncertainty relation holds for all $d$-dimensional $N$-fermion systems subject to a central potential of arbitrary type. It is interesting to note that for the values $l\equiv \mu_{1} = \ldots = m = 0$, one has 
\begin{equation}
\label{eq:ineq6}
I[\rho] \times  I[\gamma] \geq \left\{ \frac{4d}{d+1}\left[\Gamma(d+1)\right]^{\frac{1}{d}} \right\}^{2} (2 s+1)^{-\frac{2}{d}} N^{\frac{2}{d}+2},
\end{equation}
which is the Pauli-modified expression of the general spinless Fisher-information-based uncertainty relation $I[\rho] \times  I[\gamma] \geq 4d^{2} N^{2}$ recently found \cite{arplastino}. Note that for large values of the spatial dimensionality $d$, the bound (\ref{eq:ineq6}) behaves as $16 e^{-2} d^{2} N^{2} = 2.16536 \, d^{2} N^{2}$. Again here, it is manifest the delicate balance between the spatial and spin dimensionality effects which makes the lower bound (\ref{eq:ineq6}) to be better or worse than the spinless bound when $d$ is small or large, respectively. On the other hand, we observe as in the Heisenberg-like case discussed in the previous section, that the lower bound on the position-momentum Fisher-based product increases when the spatial dimen- sionality is increasing; and it decreases when the spin dimensionality is increasing, so that the Pauli effects worse the uncertainty relation, especially when the spatial dimensionality decreases. The global improvement of the Pauli-modified bound actually comes from the extra (2/d)-power which N has with respect to the spinless bound.\\
Then, for $d = 3$ we obtain the uncertainty relation
\[
I[\rho] \times  I[\gamma] \geq 9\times 6^{2/3}\, (2 s+1)^{-\frac{2}{3}} N^{\frac{8}{3}}
\]
for real $N$-fermion systems. So, for electronic systems ($s=\frac{1}{2}$), one has
\[
I[\rho] \times  I[\gamma] \geq 3^{\frac{8}{3}}N^{\frac{8}{3}}
\]
The accuracy of this relation is numerically examined in Fig. \ref{fig:2} for all neutral atoms of the periodic table from Hydrogen ($N = 1$) through Lawrencium ($N = 103$) in a Hartree-Fock framework. The lower bound is $\log 3^{8/3}=1.272\ldots$. Here again we first observe that the known atomic shell structure is grasped by the Fisher-information-based uncertainty product. Moreover, contrary to the Heisenberg uncertainty product, this Fisher-like product globally decreases, and so its accuracy globally increases, when the nuclear charge of the atom is increasing. The different behavior of the Fisher uncertainty relation with respect to the Heisenberg one is due to the local character of the position and momentum Fisher informations; indeed, they are functionals of the gradient of position and momentum densities of the system, respectively.

\section{Conclusions and open problems}

The Pauli-principle effects on two uncertainty relations (the generalised Heisenberg relations and the Fisher-information-based  relation) have been investigated together with the spatial dimensionality contribution. Summarizing, we have explicitly found the effects of the combined contribution of the spatial and spin degrees of freedom on two mathematical formulations of the quantum-mechanical uncertainty principle. First, for a system with a fixed number $N$ of fermions we have observed in Heisenberg-like and Fisher cases that the lower bound increases when both spatial and spin dimensionalities are increasing; thus the uncertainty relation becomes more accurate, so better. Second, when $N$ is increasing, the lower bound on the Heisenberg-like and Fisher-like uncertainty products globally increases and decreases, respectively; so that the corresponding uncertainty relations worse and better, respectively. The main reason for this opposite behavior is that the positIon-and-momentum-uncertainty measures of the two uncertainty relations have a global character in the Heisenberg-like case and a local character in the Fisher-like case.\\

The Pauli effects on the uncertainty relations based on the Shannon, Rényi or the Tsallis entropies remain unknown. To determine them it is necessary to design a \textit{modus operandi} different to the one used in this work. Indeed, here we have expressed the Heisenberg-like and Fisher-like uncertainty products in terms of the standard Heisenberg product $\langle r^2\rangle  \langle p^2\rangle$, and then we have obtained a lower bound on it. In the Shannon and Rényi cases it is not yet possible to express the corresponding position and momentum sums \cite{bialynicki,zozor,rudnicki} in terms of the standard Heisenberg product, and a similar situation occurs for the position and momentum quotients in the Tsallis or Rajagopal-Maassen-Uffink case \cite{rajagopal,maassen}.

\acknowledgments This work was partially supported by the Projects
FQM-2445 and FQM-207 of the Junta de Andalucia and the grant
FIS2011-24540 of the Ministerio de Innovaci\'on y Ciencia (Spain).

\end{document}